\documentclass{article}

\usepackage{arxiv}

\usepackage{tikz}
\graphicspath{{images/}}
\usepackage{amsmath}
\usepackage{latexsym, amssymb}
\usepackage{graphicx}
\usepackage{amsmath, amsbsy}
\usepackage{amsopn, amstext}
\usepackage{ifpdf,hyperref}
\usepackage{cancel, color}
\usepackage{epstopdf}

\usepackage[utf8]{inputenc} 
\usepackage[T1]{fontenc}    
\usepackage{hyperref}       
\usepackage{url}            
\usepackage{booktabs}       
\usepackage{amsfonts}       
\usepackage{nicefrac}       
\usepackage{microtype}      
\usepackage{lipsum}
\usepackage{graphicx}
\graphicspath{ {./images/} }

\def\cal{\mathcal}

\def\d{\delta}
\def\D{\Delta}

\newcommand{\R}{{\mathbb R}}

\newtheorem{thm}{Theorem}[section]
\newtheorem{dfn}[thm]{Definition}
\newtheorem{prp}[thm]{Proposition}
\newtheorem{exa}[thm]{Example}
\newtheorem{lem}[thm]{Lemma}

\newtheorem{rem}[thm]{Remark}
\newtheorem{alg}[thm]{Algorithm}

\title{Semi-Tensor Product-Based Convolutional Neural Network}

\author{
Daizhan Cheng \\
  Key Laboratory of Systems and Control\\Academy of Mathematics and Systems Sciences\\
  Chinese Academy of Sciences\\
  Beijing 100190, China\\
  \texttt{dcheng@iss.ac.cn} \\
   \And
 Xiao Zhang \\
 Department of Applied Mathematics\\
Hong Kong Polytechnic University\\
Hong Kong, 999077, China\\
 \texttt{xiaozhang@amss.ac.cn} \\
}

\begin{document}
\maketitle
\begin{abstract}
The semi-tensor product (STP) of vectors generalizes the conventional inner product, enabling algebraic operations between vectors of different dimensions. Building upon this foundation, we introduce a domain-based convolutional product (CP) and integrate it with the STP to formulate a padding-free convolutional operation. This new operation inherently avoids zero or other artificial padding, thereby eliminating redundant information and boundary artifacts commonly present in conventional convolutional neural networks (CNNs). Based on this operation, we further develop an STP-based CNN framework that extends convolutional computation to irregular and cross-dimensional data domains. Applications to image processing and third-order signal identification demonstrate the proposed method’s effectiveness in handling irregular, incomplete, and high-dimensional data without the distortions caused by padding.
\end{abstract}

\keywords{semi-tensor product \and domain-based convolution \and convolutional neural network \and irregular image \and higher-order signal identification}

\section{Introduction}
Convolutional neural networks (CNNs) have become a cornerstone of modern artificial intelligence, delivering state-of-the-art performance across domains such as computer vision, speech recognition, and natural language understanding. From the early success of gradient-based learning and large-scale image classification to the breakthroughs in residual networks and AlphaGo, convolution-based architectures have consistently demonstrated strong inductive biases for spatial and spatiotemporal data \cite{lec98,kri12,he16,goo16,sil16}. In typical vision pipelines, convolution computes local weighted aggregations over fixed-size receptive fields, enabling hierarchical feature extraction with translation equivariance. However, in finite domains and real-world datasets with boundaries, irregularities, or missing regions, conventional convolution relies on explicit padding (e.g., zero, mirror, or circular), masking, or ad hoc fixes to maintain output dimensionality and kernel alignment. These auxiliary elements are not part of the true observation and may introduce artifacts or statistical bias, thereby degrading feature purity and downstream task performance.

To mitigate boundary effects and handle data incompleteness, a wide range of operator-level enhancements have been developed. Dilated convolutions and multi-scale mechanisms expand receptive fields without increasing parameters \cite{yu16}; deformable convolutions learn spatial offsets to adapt sampling to content \cite{dai17}; and sparse or submanifold convolutions improve efficiency on irregular supports \cite{gra17}. For free-form holes and masks, partial and gated convolutions dynamically reweight valid entries to improve inpainting and reconstruction \cite{liu18,yu19}. Complementary to these algorithmic advances, matrix-based acceleration techniques (e.g., \textit{im2col}) and optimized computational primitives have enhanced deployment efficiency in industrial libraries \cite{che14,lav16}. Despite these achievements, most approaches remain fundamentally dependent on fixed-dimensional vector or tensor operations. When the number of valid samples within a receptive field varies spatially or according to data quality, the conventional inner product and tensor calculus cannot naturally accommodate cross-dimensional interactions without padding or explicit masking. Beyond practical inefficiency, the reliance on padding also reflects a theoretical inconsistency: convolutional operations are locally linear but globally constrained by fixed tensor shapes. This rigidity prevents convolution from being a truly domain-aware operator: its algebra is defined in Euclidean grid space rather than over the actual support of valid data. Consequently, once the receptive field overlaps incomplete or irregular regions, conventional CNNs lose their algebraic consistency, forcing heuristic treatments such as masking or pseudo-padding.

Motivated by these limitations, this paper introduces a new convolutional framework grounded in the semi-tensor product (STP) of vectors, which enables algebraically consistent operations between vectors of different dimensions \cite{che19,che19b,che24}. While deformable and sparse convolutions relax spatial constraints, they still operate within fixed-dimensional tensor spaces. In contrast, the proposed STP-based convolution operates in $\mathbb{R}^\infty$, providing a principled, dimension-free formulation that inherently accommodates variable receptive field sizes without the need for explicit padding or masking. From a mathematical viewpoint, the STP provides a natural bridge between continuous-domain algebra and discrete tensor computation. It extends the conventional tensor inner product to a cross-dimensional setting, preserving linearity and associativity while removing the dependency on fixed array dimensions. This property makes it uniquely suitable for modeling convolution over irregular, incomplete, or multi-resolution domains, where the notion of a fixed grid is no longer adequate.

To preserve both theoretical rigor and engineering practicality, we employ a block Hadamard product and a receptive field matrix (RFM) to express the sliding-window computation in a unified linear-algebraic form. This construction yields an STP-based CNN (STP-CNN) whose convolutional operation is defined by the inner product between variable-length valid token vectors and fixed kernels. The resulting framework generalizes naturally to irregular images, partially damaged inputs, and proportional convolution where receptive field and kernel sizes are mismatched. Furthermore, the method is extended to third-order signals (e.g., 3D medical imaging) via depth-wise masking and rearrangement, enabling effective processing of volumetric and multi-channel data \cite{kil13,lim13,cic16}.

The main contributions of this work are as follows:
\begin{itemize}
\item {\bf Padding-free convolution operator:} We propose a convolution operator based on the STP that computes cross-dimensional inner products solely over valid data. This design eliminates zero-padding and explicit mask propagation, thereby reducing boundary artifacts and enhancing robustness to irregular or partially missing inputs.
\item {\bf Unified linear-algebraic formulation:} A concise formulation based on the block Hadamard product and RFM expresses the STP convolution in matrix form, maintaining compatibility with mainstream deep learning implementations while grounded in rigorous STP theory.
\item {\bf Scalable generalization to 3D and mismatched scales:} The framework naturally supports proportional convolution (where receptive field and kernel sizes differ) and extends to volumetric data via depth-wise masking and rearrangement, unifying 2D irregular images, damaged inputs, and 3D signals within a single paradigm.
\end{itemize}

In summary, we propose an STP-CNN that replaces padding-dependent inner products with cross-dimensional STP operations and domain-restricted convolution. This approach preserves the mathematical essence of convolution while minimizing the introduction of extraneous information, demonstrating strong adaptability across 2D irregular images, partially damaged inputs, and 3D volumetric signals. We believe this direction offers a rigorous and practical pathway toward representation learning on high-dimensional, irregular, and incomplete data. Future work will focus on large-scale empirical evaluation, efficient integration with mainstream libraries \cite{che14,lav16}, and systematic comparisons with deformable, sparse, and gated convolutional baselines.

The remainder of this paper is organized as follows:
Section~\ref{S2} reviews the STP and the cross-dimensional inner product on $\mathbb{R}^\infty$ as the algebraic foundation.
Section~\ref{S3} defines the domain-based convolutional product (DCP) and establishes its properties.
Section~\ref{S4} introduces the block Hadamard product and RFM.
Section~\ref{S5} formulates the STP-based convolutional product (STP-CP) operating without padding.
Section~\ref{S6} demonstrates STP-CP on irregular and partially damaged images as well as proportional convolution.
Section~\ref{S7} extends the framework to 3D signals with depth masking and rearrangement.
Section~\ref{S8} concludes and outlines directions for large-scale experimentation and library integration.

\section{Semi-tensor product of vectors\label{S2}}
To establish the algebraic foundation of the proposed framework, this section reviews the STP of vectors and the associated cross-dimensional inner product defined on $\mathbb{R}^\infty$. The STP enables consistent operations between vectors of different dimensions, overcoming the fixed-size limitation of conventional inner products. We further introduce the corresponding norm, distance, and equivalence relations, which together form a topological vector space structure that will serve as the mathematical basis for subsequent developments.

Consider the cross-dimensional Euclidean space:
\begin{align*} \mathbb{R}^{\infty} := \bigcup_{n=1}^{\infty} \mathbb{R}^n. \end{align*}

\begin{dfn}\label{d2.1}
Let $x \in \mathbb{R}^m \subset \mathbb{R}^{\infty}$ and $y \in \mathbb{R}^n \subset \mathbb{R}^{\infty}$, with $t = \mathrm{lcm}(m,n)$.
The addition (or subtraction) of $x$ and $y$ is defined as
\begin{align}\label{2.1}
x \vec{\pm} y := (x \otimes {\bf 1}_{t/m}) \pm (y \otimes {\bf 1}_{t/n}) \in \mathbb{R}^t,
\end{align}
where $\otimes$ denotes the Kronecker product and ${\bf 1}_k$ is a column vector of ones of length $k$.
\end{dfn}

\begin{dfn}\label{d2.2}
Let $x \in \mathbb{R}^m \subset \mathbb{R}^{\infty}$ and $y \in \mathbb{R}^n \subset \mathbb{R}^{\infty}$.
\begin{itemize}
\item[(i)] \textbf{Inner product:}
\begin{align}\label{2.2}
\langle x, y \rangle_{\mathcal{V}} :=
\frac{1}{t} \langle (x \otimes {\bf 1}_{t/m}), (y \otimes {\bf 1}_{t/n}) \rangle,
\end{align}
where $t = \mathrm{lcm}(m,n)$ and $\langle \cdot,\cdot \rangle$ is the standard Euclidean inner product.
\item[(ii)] \textbf{Norm:}
\begin{align}\label{2.3}
\|x\|_{\mathcal{V}} := \sqrt{\langle x, x \rangle_{\mathcal{V}}}.
\end{align}
\item[(iii)] \textbf{Distance:}
\begin{align}\label{2.4}
d_{\mathcal{V}}(x, y) := \|x \vec{-} y\|_{\mathcal{V}}.
\end{align}
\end{itemize}
\end{dfn}

\begin{lem}[\cite{che19}]\label{p2.3}
\begin{itemize}
\item[(i)] Under the addition defined in \eqref{2.1} and the conventional scalar multiplication, $\mathbb{R}^{\infty}$ forms a pseudo-vector space.
\item[(ii)] Equipped with the topology induced by the distance in \eqref{2.4}, $\mathbb{R}^{\infty}$ is a topological space.
\end{itemize}
\end{lem}

\begin{dfn}\label{d2.4}
For $x, y \in \mathbb{R}^{\infty}$, we say $x$ and $y$ are \emph{equivalent}, denoted by $x \leftrightarrow y$, if
\begin{align*}
d_{\mathcal{V}}(x, y) = 0.
\end{align*}
The equivalence class of $x$ is then defined as
\begin{align}\label{2.5}
\bar{x} := \{\, y \mid y \leftrightarrow x \,\}.
\end{align}
\end{dfn}

\begin{dfn}\label{d2.5}
Let $\Omega$ denote the set of equivalence classes:
\begin{align*}
\Omega = \mathbb{R}^{\infty} / \!\leftrightarrow := \{\, \bar{x} \mid x \in \mathbb{R}^{\infty} \,\}.
\end{align*}
For $\bar{x},\bar{y}\in\Omega$, we define
\begin{itemize}
\item[(i)] \textbf{Addition:}
\begin{align}\label{2.6}
\bar{x} \vec{\pm} \bar{y} := x \vec{\pm} y, \qquad \bar{x}, \bar{y} \in \Omega.
\end{align}

\item[(ii)] \textbf{Inner product:}
\begin{align}\label{2.7}
\langle \bar{x}, \bar{y} \rangle_{\mathcal{V}} := \langle x, y \rangle_{\mathcal{V}}, \qquad \bar{x}, \bar{y} \in \Omega.
\end{align}

\item[(iii)] \textbf{Norm:}
\begin{align}\label{2.8}
\|\bar{x}\|_{\mathcal{V}} := \|x\|_{\mathcal{V}}.
\end{align}

\item[(iv)] \textbf{Distance:}
\begin{align}\label{2.9}
d_{\mathcal{V}}(\bar{x}, \bar{y}) := d_{\mathcal{V}}(x, y).
\end{align}
\end{itemize}
\end{dfn}

\begin{lem}[\cite{che24}]\label{p2.6}
\begin{itemize} \item[(i)] The definitions in \eqref{2.6}–\eqref{2.9} are well-defined, i.e., they are independent of the choice of representative elements.
\item[(ii)] Under these operations and the topology induced by the distance in \eqref{2.9}, $\Omega$ forms a topological vector space (see \cite{sch70} for details). \end{itemize}
\end{lem}

\begin{figure}[h]
  \centering
  \includegraphics{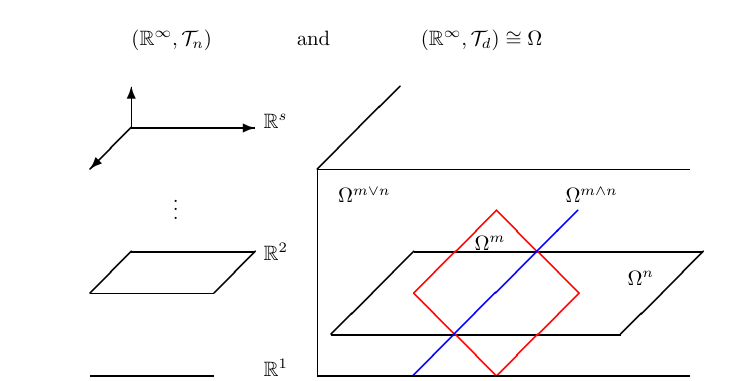}
  \caption{$(\R^{\infty},{\cal T}_n)$ vs~$\Omega$}
  \label{Fig.2.01}
\end{figure}

Fig. \ref{Fig.2.01} illustrates the geometric structure of $\Omega$ in comparison to the conventional Euclidean structure, where ${\cal T}_n$ denotes the natural topology and ${\cal T}_d$ represents the topology induced by the distance.

\section{Classical CP and domain-based CP\label{S3}}
Building on the preliminaries of the STP, this section revisits the classical convolutional product (CP) in both continuous and discrete forms and extends it to DCP. Unlike the traditional CP, DCP restricts the integration or summation to the valid data region, which naturally aligns with real-world cases where signals or images contain missing or irregular portions. This section also summarizes several key properties of DCP, preparing the groundwork for its matrix-form implementation in the next section.

\subsection{Classical CP}

\begin{dfn}\label{d3.1.1}
Let $f(x), g(x) \in L^1(\mathbb{R})$.
The CP of $f(x)$ and $g(x)$, denoted by $f(x) * g(x)$, is defined as
\begin{align}\label{3.1.1}
s(x) := f(x) * g(x) = \int_{-\infty}^{\infty} f(\tau)\, g(x - \tau)\, d\tau.
\end{align}
\end{dfn}

Some fundamental properties of CP are summarized below.

\begin{prp}[\cite{edw79}]\label{p3.1.2}
\begin{itemize} Key properties of the CP include:
\item[(i)] $L^1$-Stability:
\begin{align}\label{3.1.2}
\|f(x) * g(x)\|_1 \leq \|f(x)\|_1 \, \|g(x)\|_1.
\end{align}
Hence, $f(x) * g(x) \in L^1(\mathbb{R})$.
\item[(ii)] Commutativity:
\begin{align}\label{3.1.3}
f(x) * g(x) = g(x) * f(x).
\end{align}
\item[(iii)] Associativity:
If $f(x), g(x), h(x) \in L^1(\mathbb{R})$, then
\begin{align}\label{3.1.4}
f(x) * (g(x) * h(x)) = (f(x) * g(x)) * h(x).
\end{align}
\end{itemize}
\end{prp}

The above properties concern the continuous setting. For discrete-time signals, convolution is defined as follows.

\begin{dfn}\label{d3.1.3}
Let $f(n), w(n) \in \ell^1$. The discrete-time CP of $f(n)$ and $w(n)$ is
\begin{align}\label{3.1.5}
s(n) := f(n) * w(n) = \sum_{\tau = -\infty}^{\infty} f(\tau)\, w(n - \tau).
\end{align}
\end{dfn}

When $w$ is regarded as a weighting function, \eqref{3.1.1} and \eqref{3.1.5} can be interpreted as a weighted average,
where the weights depend on the distance between the current position and the arguments of the function being averaged.
Proposition~\ref{p3.1.2} also holds for the discrete-time CP.
Using \eqref{3.1.2}, we further obtain the following result.

\begin{prp}\label{p3.1.4}
The stability results of Proposition \ref{p3.1.2} extend directly to the discrete-time case and finite windows:
\begin{itemize}
\item[(i)] If $f(x), w(x) \in L^1$, then $f(x) * w(x) \in L^1$.
\item[(ii)] If $f(n), w(n) \in \ell^1$, then $f(n) * w(n) \in \ell^1$.
\end{itemize}
\end{prp}

\subsection{CP with finite windows}

In deep learning and many signal processing tasks, convolution is computed over a finite window:
\begin{align}\label{3.2.1}
s(n) := f(n) * k(n) = \sum_{\tau = -N}^{N} f(\tau)\, k(n - \tau),
\end{align}
where $[-N, N]$ is the \emph{sampling window}, $f(n)$ is the input signal, and $k(n)$ is the convolution kernel.

By exploiting the commutativity property in \eqref{3.1.3}, one may alternatively write
\begin{align}\label{3.2.2}
s(n) := f(n) * k(n) = \sum_{\tau = -N}^{N} f(n - \tau)\, k(\tau).
\end{align}

In convolutional neural networks (CNNs), a closely related operation—\emph{cross-correlation}—is often used, typically defined as
\begin{align}\nonumber
s(n) :=& f(n) * k(n)\\\label{3.2.3}
=& \sum_{\tau = -N}^{N} f(n + \tau)\, k(\tau).
\end{align}
Note: although the same symbol ``$*$" is often used in practice, \eqref{3.2.3} corresponds to cross-correlation rather than the commutative convolution in \eqref{3.2.1}-\eqref{3.2.2}.

For multivariate functions, CP generalizes naturally.
For two-variable functions,
\begin{align}\nonumber
s(m,n) :=& f(m,n) * k(m,n) \\\label{3.2.4}
=& \sum_{i} \sum_{j} f(i,j)\, k(m - i, n - j),
\end{align}
or equivalently,
\begin{align}\nonumber
s(m,n) :=& f(m,n) * k(m,n) \\\label{3.2.5}
=& \sum_{i} \sum_{j} f(m - i, n - j)\, k(i,j).
\end{align}
This two-dimensional CP is standard in image processing.

\subsection{Domain-based CP}

\begin{dfn}\label{d3.3.1}
Let $D \subset \mathbb{R}$ be measurable, $f(x) \in L^1(D)$, and $w(x) \in L^1(\mathbb{R})$.
The \emph{domain-based convolutional product (DCP)} of $f(x)$ with $w(x)$ is
\begin{align}\label{3.3.1}
s(x) := f(x) * w(x) = \int_{\tau \in D} f(\tau)\, w(x - \tau)\, d\tau.
\end{align}
\end{dfn}

Several basic properties of classical CP extend directly to DCP.

\begin{prp}\label{p3.3.2}
\begin{itemize}
\item[(i)] The domain-based convolution $f(x)*w(x)$ is a measurable function on $\mathbb{R}$. Moreover, we have \begin{align}\label{3.3.2}
\|f(x) * w(x)\|_1 \leq \|f(x)\|^{D}_1 \, \|w(x)\|_1<\infty.
\end{align} Therefore, $f(x) * w(x) \in L^1(\mathbb{R})$.
\item[(ii)]
The commutativity relation holds in the following sense:
\begin{align}\label{3.3.3}
\int_{\tau \in D} f(\tau)\, w(t - \tau)\, d\tau
= \int_{t - \tau \in D} w(\tau)\, f(t - \tau)\, d\tau.
\end{align}
\item[(iii)]
If $f(x) \in L^1(D)$ and $g(x), h(x) \in L^1(\mathbb{R})$, then
\begin{align}\label{3.3.4}
f(x) * (g(x) * h(x)) = (f(x) * g(x)) * h(x).
\end{align}
\end{itemize}
\end{prp}
\begin{rem}
    The proofs of Proposition \ref{p3.3.2} follow straightforwardly from Definition \ref{d3.3.1} and the properties of the classical convolution. They are therefore omitted.
\end{rem}
\section{CP for image processing\label{S4}}
This section develops a matrix-based representation of classical CNN convolution, establishing a unified algebraic framework for the subsequent derivation. We first recall the standard procedure of image convolution with zero-padding and then introduce two key matrix constructs: the block Hadamard product and the RFM. In classical CNNs, the receptive field is an implicit, sliding-window concept referring to the local input region (e.g., an $s \times t$ patch) processed by the kernel. The RFM $R_A$ provides an explicit matrix representation by gathering all receptive fields of the padded image $\bar{A}$ into a structured block matrix. These tools not only clarify the linear-algebraic essence of convolution but also pave the way for its extension to STP-based operations in Section \ref{S5}.

\subsection{Classical CP in CNN for image processing}

The CP  used in CNNs for image processing is well documented in standard references (see, e.g., \cite{goo16,zha23,wan24}).

Given an image matrix $A \in \mathbb{R}^{m \times n}$ and a kernel matrix $K \in \mathbb{R}^{s \times t}$, the convolution proceeds as follows.

\begin{alg}\label{asc.2.1.1}
\begin{itemize}
\item \textbf{Step 1:}  \textbf{Padding.}
To address boundary effects, $A$ is padded (typically with zeros) so that all original pixels become interior points. Let the padded matrix be $\bar{A} \in \mathbb{R}^{p \times q}$, where
\begin{align}\label{sc.2.1.1}
p = m + 2\Delta_v, \qquad q = n + 2\Delta_h,
\end{align}
and
\begin{align}\label{sc.2.1.2}
\bar{a}_{i,j} =
\begin{cases}
a_{i - \Delta_v, j - \Delta_h}, & \Delta_v + 1 \le i \le \Delta_v + m,\\& \Delta_h + 1 \le j \le \Delta_h + n, \\
0, & \text{otherwise}.
\end{cases}
\end{align}
This process is illustrated in Fig.~\ref{Fig.2.1}.
\begin{figure}[h]
  \centering
  \includegraphics{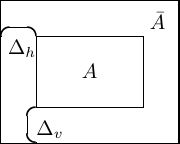}
  \caption{Illustration of image padding to obtain $\bar{A}$.\label{Fig.2.1}}
\end{figure}

\item \textbf{Step 2:}  \textbf{Local convolution.}
For each receptive field $\bar{A}_{i,j}$ of $\bar{A}$ covered by the kernel $K$, compute
\begin{align}\label{sc.2.1.3}
s_{i,j} = \langle V_c(K), V_c(\bar{A}_{i,j}) \rangle,
\end{align}
where $V_c(\cdot)$ denotes the column stacking form of a matrix.
\item \textbf{Step 3:}  \textbf{Sliding and output construction.}
Slide $K$ across $\bar{A}$ with vertical and horizontal strides $d_v$ and $d_h$, respectively, to form $S = (s_{i,j}) \in \mathbb{R}^{s_v \times s_h}$ satisfying
\begin{align}\label{sc.2.1.4}
\begin{cases}
(s_v - 1)d_v + s = p,\\
(s_h - 1)d_h + t = q.
\end{cases}
\end{align}
The padding sizes $\Delta_v$ and $\Delta_h$ are chosen so that \eqref{sc.2.1.4} holds.
\end{itemize}
\end{alg}

\begin{figure}[h]
  \centering
  \includegraphics{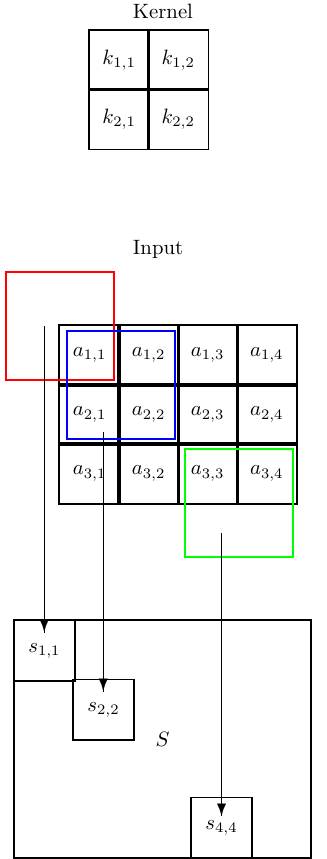}
  \caption{Image CP\label{Figsc.2.1}}
\end{figure}

The following example demonstrates the algorithm with a concrete case.
\begin{exa}\label{esc.2.1.2}
Consider a grayscale image $A \in \mathbb{R}^{3 \times 4}$ and a kernel $K \in \mathbb{R}^{2 \times 2}$, where
\[
A =
\begin{bmatrix}
1 & 2 & -1 & -2\\
-3 & -2 & 1 & 3\\
2 & -2 & 1 & -1
\end{bmatrix},
\quad
K =
\begin{bmatrix}
1 & 0.4\\
0.6 & 1.5
\end{bmatrix}.
\]
With $\Delta_v = \Delta_h = 1$, the padded image is $\bar{A} \in \mathbb{R}^{5 \times 6}$.
Using \eqref{sc.2.1.3}, each $s_{i,j}$ is computed by
\begin{align}\label{sc.2.1.5}
s_{s,t} = \sum_{i=1}^2 \sum_{j=1}^2 \bar{a}_{s - 2 + i, t - 2 + j} \, k_{i,j},
\quad 2 \le s \le 4,~  2\le t \le 5.
\end{align}
The resulting CP matrix is
\[
S =
\begin{bmatrix}
1.5 & 3.6 & -0.3 & -3.6 & -1.2\\
-4.1 & -3.0 & 1.9 & 3.2 & -0.2\\
1.8 & -5.6 & -1.3 & 1.3 & 2.4\\
0.8 & 1.2 & -1.6 & 0.6 & -1
\end{bmatrix}.
\]

Fig. \ref{Figsc.2.1} shows a schematic of the input, kernel, and selected outputs (e.g., $s_{1,1}, s_{2,2}, s_{4,4}$).
\end{exa}

\subsection{Block Hadamard product and receptive field matrix}

To compute CP efficiently, we introduce two matrix structures:
the \emph{block Hadamard product} and the \emph{receptive field matrix (RFM)}.

\begin{dfn}[Block Hadamard Product]\label{dsc.2.101.1}
Let $A$ and $B$ be two matrices of identical size and block structure:
\begin{align}\label{sc.2.101.1}
A =
\begin{bmatrix}
A^{1,1} & \cdots & A^{1,q}\\
\vdots & \ddots & \vdots\\
A^{p,1} & \cdots & A^{p,q}
\end{bmatrix},
\quad
B =
\begin{bmatrix}
B^{1,1} & \cdots & B^{1,q}\\
\vdots & \ddots & \vdots\\
B^{p,1} & \cdots & B^{p,q}
\end{bmatrix}.
\end{align}
The \emph{block Hadamard product} of $A$ and $B$, denoted by $A \vec{\circ} B$, is defined as
\begin{align}\label{sc.2.101.2}
A \vec{\circ} B := C = (c_{i,j}) \in \mathbb{R}^{p \times q},
\end{align}
where
\[
c_{i,j} = \langle V_c(A^{i,j}), V_c(B^{i,j}) \rangle, \quad i \in [1,p],~ j \in [1,q].
\]
\end{dfn}

\begin{figure}[h]
  \centering
  \includegraphics{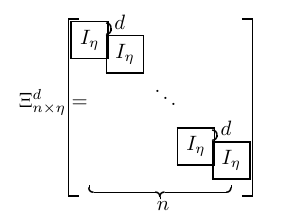}
  \caption{Matrix $\Xi_{n\times \eta}^d$\label{Figsc.2.2}}
\end{figure}

\begin{dfn}[Receptive Field Matrix (RFM)]\label{dsc.2.2.1}
Let $\bar{A} \in \mathbb{R}^{p \times q}$ be a padded image satisfying \eqref{sc.2.1.4}.
Define the RFM $R_A$ as
\begin{align}\label{sc.2.2.2}
R_A = P \bar{A} Q \in \mathbb{R}^{(s_v s) \times (s_h t)},
\end{align}
where
\begin{align}\label{sc.2.2.3}
P^{\mathrm{T}} = \Xi_{s_v \times s}^{d_v}, \quad Q = \Xi_{s_h \times t}^{d_h}.
\end{align}
Here, $\Xi_{n \times \eta}^d\in \mathbb{R}^{\xi \times (n\eta)}$ is a structured selection matrix with stride $d$, $\xi = (n-1)d + \eta$, as illustrated in Fig.~\ref{Figsc.2.2}.
\end{dfn}

The matrix $R_A$ admits a block form:
\begin{align}\label{sc.2.2.4}
R_A =
\begin{bmatrix}
R_{1,1} & \cdots & R_{1,s_h}\\
\vdots & \ddots & \vdots\\
R_{s_v,1} & \cdots & R_{s_v,s_h}
\end{bmatrix},
\end{align}
where each $R_{i,j} \in \mathbb{R}^{s \times t}$ corresponds to a receptive field of $\bar{A}$. Then the CP matrix entries are
\begin{align}\label{sc.2.2.5}
s_{i,j} = \langle V_c(K), V_c(R_{i,j}) \rangle, \quad
i \in [1,s_v],~ j \in [1,s_h].
\end{align}

Using the block Hadamard product, the entire convolutional output can be compactly expressed as follows.

\begin{prp}\label{psc.2.2.2}
Let $R_A$ be defined as in \eqref{sc.2.2.4}, where each block $R_{i,j}$ matches the size of $K$. Then the convolutional output matrix is
\begin{align}\label{sc.2.2.6}
S = ({\bf 1}_{s_v \times s_h} \otimes K) \vec{\circ} R_A.
\end{align}
\end{prp}

{\it Proof.} This follows directly from the construction of the RFM $R_A$
  in \eqref{sc.2.2.4} and the definition of the block Hadamard product in \eqref{sc.2.101.2}, which together implement the sliding-window inner product described in Algorithm \ref{asc.2.1.1}.
  \hfill $\Box$

\section{Semi-tensor product-based CP\label{S5}}
With the mathematical and structural groundwork in place, this section formally defines the STP-CP. The proposed operation replaces the conventional inner product with the STP-based cross-dimensional inner product, allowing convolution to be performed directly on valid entries without zero-padding or explicit mask propagation. Through examples and remarks, we illustrate how STP-CP generalizes convolution to variable-sized receptive fields, enabling a padding-free and dimension-flexible CNN formulation.

We recall the STP of vectors defined in \eqref{2.2} for completeness. In CP of CNNs, each element of the CP matrix $S$ is computed via a conventional inner product, as in \eqref{sc.2.1.3}.

In the proposed STP-CP, consider a receptive field $\bar{A}_{i,j}$.
If certain entries in this receptive field are undefined (e.g., at image boundaries or within damaged regions), we do not pad them with zeros; instead, we simply ignore the invalid entries. Consequently, different receptive fields may have different effective dimensions, and the STP of vectors of varying lengths is used to compute the inner product between $K$ and $A_{i,j}$.
This yields the \emph{STP-CNN}, a convolutional neural network constructed using the STP-CP.
Formally, it is defined as follows.

\begin{dfn}\label{dsc.3.1}
Let $A$ be the image matrix and $K$ the kernel matrix.
The STP-based convolutional product is:
\begin{align}\label{sc.3.2}
s_{i,j} := \langle V_c(A_{i,j}), V_c(K) \rangle_{\mathcal{V}},
\end{align}
where $A_{i,j}$ denotes the subset of valid (available) entries within the receptive field corresponding to position $(i,j)$.
\end{dfn}

\begin{rem}\label{rsc.3.2}

Referring to Fig.~\ref{Figsc.2.1}, consider several receptive fields of varying dimensions: the field for $s_{1,1}$ is a $1 \times 1$ matrix $A_{1,1} \in \mathbb{R}^{1 \times 1}$, that for $s_{2,2}$ is $A_{2,2} \in \mathbb{R}^{2 \times 2}$, and for $s_{4,4}$ it is $A_{4,4} \in \mathbb{R}^{1 \times 2}$.
Because these receptive fields have different dimensions, their corresponding inner products with $K$ are computed via the STP of vectors of different sizes.

\end{rem}

\begin{exa}\label{esc.3.3}
Continuing Example~\ref{esc.2.1.2}, applying the STP-CP as in \eqref{sc.3.2} yields:
\[
S = \frac{1}{4}
\begin{bmatrix}
3.5 & 5.4 & 1.3 & -5.4 & -7.0\\
-4.1 & -3 & 1.9 & 3.3 & 2.5\\
-1& -5.6 & -1.3 & 1.3 & 2.9\\
7.0 & -0.4 & -0.7 & -0.7 & -3.5
\end{bmatrix}.
\]
\end{exa}

\begin{rem}{\bf Computational Complexity of STP-CP} To provide a formal comparison, we analyze the computational profile of the proposed STP-CP operation, which generalizes classical convolution. Classical convolution corresponds to the special case of STP-CP where all receptive fields are normal (i.e., of the full $s \times t$ size).

The complexity of classical CP for a single output element with an $s \times t$ kernel is $O(C_{\text{in}} \cdot s \cdot t)$. The overall complexity for an entire layer is $O(O_h \cdot O_w \cdot C_{\text{in}} \cdot C_{\text{out}} \cdot s \cdot t)$, where $O_h \times O_w$ is the output size, $C_{\text{in}}$ is the number of input channels, and $C_{\text{out}}$ is the number of output channels (i.e., filters).

For STP-CP, we consider two scenarios regarding the receptive fields (the local input regions processed by the kernel):
\begin{itemize}
    \item \textbf{Normal (Full-sized) Receptive Fields:} When the receptive field is complete, STP-CP is computationally identical to classical CP for a single output element, i.e., $O(C_{\text{in}} \cdot s \cdot t)$.

    \item \textbf{Abnormal (Irregular) Receptive Fields:} For a receptive field with $v$ valid entries ($v < s \cdot t$), STP-CP incurs an overhead, with a per-element complexity of $O(C_{\text{in}} \cdot \ell)$, where $\ell = \mathrm{lcm}(v, s \cdot t)$. This overhead stems from calculating $\ell$ and performing the requisite Kronecker products.
\end{itemize}

Crucially, since the number of abnormal receptive fields is typically small (e.g., confined to boundaries or sparse damaged regions), the \textbf{overall additional cost is amortized over the entire input} and does not alter the asymptotic complexity of the layer. Furthermore, the \textbf{parameter count of an STP-CNN layer remains exactly the same} as its standard CNN counterpart.
\end{rem}

\section{Application of STP-CP\label{S6}}
This section demonstrates the practical capabilities of the proposed STP-CP framework through several representative examples. We first show how STP-CP operates effectively on irregular and partly damaged images by directly using available data, followed by an application to proportional STP-CP, which handles mismatched receptive field and kernel sizes. These examples collectively verify that STP-CP not only eliminates artifacts introduced by padding but also improves robustness to data irregularity and scale variation.
\subsection{Irregular image}

The STP-CP framework naturally handles \emph{irregular images} (those with undefined or missing pixels) by operating directly on valid entries without padding or manual repair.

\begin{exa}\label{esc.4.1.1}
Consider the irregular image in Fig.~\ref{Figsc.3.1}, with receptive fields indicated by blue rectangles.

\begin{figure}[h]
  \centering
  \includegraphics{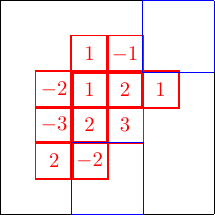}
  \caption{An irregular image and its receptive fields.\label{Figsc.3.1}}
\end{figure}

Assume the convolution kernel $K$ is the same as in Fig.~\ref{Figsc.2.1}, and let the vertical and horizontal paddings and strides satisfy
\[
\Delta_v = \Delta_h = d_v = d_h = 1.
\]
Then,
\[
\begin{array}{l}
P^{\mathrm{T}} = Q = \Xi^1_{5 \times 2} =
\begin{bmatrix}
1 & 0 & 0 & 0 & 0 & 0 & 0 & 0 & 0 & 0\\
0 & 1 & 1 & 0 & 0 & 0 & 0 & 0 & 0 & 0\\
0 & 0 & 0 & 1 & 1 & 0 & 0 & 0 & 0 & 0\\
0 & 0 & 0 & 0 & 0 & 1 & 1 & 0 & 0 & 0\\
0 & 0 & 0 & 0 & 0 & 0 & 0 & 1 & 1 & 0\\
0 & 0 & 0 & 0 & 0 & 0 & 0 & 0 & 0 & 1
\end{bmatrix}.
\end{array}
\]
Using \eqref{sc.2.2.2}, the extended image matrix is represented as
\[
M_{\bar{A}} = P \bar{A} Q := (\bar{A}^{i,j}) \in \mathbb{R}^{10 \times 10},
\]
where
\[
\bar{A}^{i,j} \in \mathbb{R}^{2 \times 2}, \quad i,j \in [1,5].
\]

Let $V_{i,j}$ denote the valid (non-missing) entries of $\bar{A}^{i,j}$.
The matrix of valid regions is
\[
V = (V_{i,j}) =
\left[
\begin{array}{ccccc}
\times & 1 & \begin{bmatrix} 1\\ -1 \end{bmatrix} & -1 & \times\\[4pt]
-2 & \begin{bmatrix} -2\\ 1\\ 1 \end{bmatrix} & \begin{bmatrix} 1\\ 1\\ -1\\ 2 \end{bmatrix} & \begin{bmatrix} -1\\ 2\\ 1 \end{bmatrix} & 1\\[4pt]
\begin{bmatrix} -2\\ -3 \end{bmatrix} & \begin{bmatrix} -2\\ -3\\ 1\\ 2 \end{bmatrix} & \begin{bmatrix} 1\\ 2\\ 2\\ 3 \end{bmatrix} &
\begin{bmatrix} 2\\ 3\\ 1 \end{bmatrix} & 1\\[4pt]
\begin{bmatrix} -3\\ 2 \end{bmatrix} & \begin{bmatrix} -3\\ 2\\ 2\\ -2 \end{bmatrix} & \begin{bmatrix} 2\\ -2\\ 3 \end{bmatrix} &
3 & \times\\[4pt]
2 & \begin{bmatrix} 2\\ -2 \end{bmatrix} & -2 & \times & \times
\end{array}
\right],
\]
where  ``$\times$’’ denotes undefined entries.

Applying the STP-CP to these irregular receptive fields yields
\begin{align}\nonumber
S &= ({\bf 1}_{5\times5} \otimes K) \vec{\circ} V\\\label{sc.3.1.1}
&= \frac{1}{12}
\begin{bmatrix}
\times & 10.5 & -0.9 & -10.5 & \times\\
-21.0 & -0.3 & 12.6 & 5.3 & 10.5\\
-26.7 & -1.2 & 22.5 & 18.1 & 10.5\\
-3.0 & -12.0 & 17.9 & 31.5 & \times\\
21.0 & -1.8 & -21.0 & \times & \times
\end{bmatrix}.
\end{align}
\end{exa}

This example demonstrates that STP-CP can perform convolution directly on irregular or incomplete image data without the need for zero-padding.
By dynamically adjusting the inner product dimensionality through the STP of vectors, the proposed method naturally adapts to the valid regions of the image, preserving structural information and avoiding artificial noise introduced by padding.

\subsection{Partly damaged image}

STP-CP also applies to partly damaged images by restricting computation to available pixels within each receptive field.

\begin{exa}\label{esc.4.2.1}
Consider the partly damaged image in Fig. \ref{Figsc.3.2}, where the damaged region is indicated by ``$\times$".

\begin{figure}[h]
  \centering
  \includegraphics{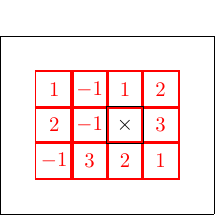}
  \caption{STP-CP applied to a partly damaged image\label{Figsc.3.2}}
\end{figure}

Assume the kernel $K$ is the same as in Fig. \ref{Figsc.2.1}. Omitting intermediate steps, the vector forms of all receptive fields are
$$
\begin{array}{l}
V=(V_{i,j})=
\left[
\begin{array}{ccccc}
1 & \begin{bmatrix}1\\-1\end{bmatrix} & \begin{bmatrix}-1\\1\end{bmatrix} & \begin{bmatrix}1\\2\end{bmatrix} & 2\\
\begin{bmatrix}1\\2\end{bmatrix} & \begin{bmatrix}1\\2\\-1\\-1\end{bmatrix} & \begin{bmatrix}-1\\-1\\1\end{bmatrix} & \begin{bmatrix}1\\2\\3\end{bmatrix} & \begin{bmatrix}2\\3\end{bmatrix}\\
\begin{bmatrix}2\\-1\end{bmatrix} & \begin{bmatrix}2\\-1\\-1\\3\end{bmatrix} & \begin{bmatrix}-1\\3\\2\end{bmatrix} & \begin{bmatrix}2\\3\\1\end{bmatrix} & \begin{bmatrix}3\\1\end{bmatrix}\\
-1 & \begin{bmatrix}-1\\3\end{bmatrix} & \begin{bmatrix}3\\2\end{bmatrix} & \begin{bmatrix}2\\1\end{bmatrix} & 1\\
\end{array}
\right].
\end{array}
$$

Applying STP-CP to the above, the resulting CP matrix is
\begin{align}\label{sc.3.2.1}
S=\frac{1}{12}\begin{bmatrix}
10.5 & -0.9 & 0.9 & 16.2 & 21\\
16.2 & 0.9 & -0.7 & 22.3 & 26.7\\
3.9 & 16.5 & 12.2 & 18.1 & 20.1\\
-10.5 & 12.3 & 25.8 & 15.3 & 10.5\\
\end{bmatrix}.
\end{align}

\end{exa}

\subsection{Proportional STP-CP}

As mentioned in \cite{wan24}, ``Assume the size of a dog is the same as the training dogs, CNN can recognize it well. But if the dog image is considerably enlarged, CNN may fail to recognize it." Conventional CNNs often struggle with large-scale variations: a model trained on objects of one size may fail when objects are significantly larger or smaller. STP-CP addresses this by allowing receptive fields whose sizes do not match the kernel size exactly. By setting receptive field sizes appropriately, a proportional STP-CNN can recognize objects across scales. The following example illustrates proportional STP-CP for mismatched image sizes.

\begin{exa}\label{esc.3.3.1}
Consider the enlarged image shown in Fig. \ref{Figsc.3.3}. Use a receptive field of size $3\times 3$ and a step length of $2$; diagonal receptive fields are highlighted in blue.

\begin{figure}[h]
  \centering
  \includegraphics{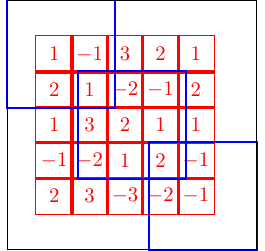}
  \caption{Proportional STP-CP for mismatched sizes\label{Figsc.3.3}}
\end{figure}

With the same kernel $K$ as in Fig. \ref{Figsc.2.1}, the vector forms of all receptive fields are
$$
\begin{array}{l}
V=(V_{i,j})=
\left[
\begin{array}{ccc}
\begin{bmatrix}1\\2\\-1\\1\end{bmatrix} & \begin{bmatrix}-1\\1\\3\\-2\\2\\-1\end{bmatrix} & \begin{bmatrix}2\\-1\\1\\2\end{bmatrix} \\
\begin{bmatrix}2\\1\\-1\\1\\3\\-2\end{bmatrix} & \begin{bmatrix}1\\3\\-2\\-2\\2\\1\\-1\\1\\2\end{bmatrix} & \begin{bmatrix}-1\\1\\2\\2\\1\\-1\end{bmatrix} \\
\begin{bmatrix}-1\\2\\-2\\3\end{bmatrix} & \begin{bmatrix}-2\\3\\1\\-3\\2\\-2\end{bmatrix} & \begin{bmatrix}2\\-2\\-1\\-1\end{bmatrix}
\end{array}
\right].
\end{array}
$$
Applying STP-CP yields
\begin{align}\label{sc.3.3.1}
S=\frac{1}{36}\begin{bmatrix}
29.7 & 7.2 & 43.2\\
14.7 & 26.5 & -9.9\\
35.1 & -7.8 & -7.8\\
\end{bmatrix}.
\end{align}

\end{exa}

\begin{rem}\label{rsc.3.3.2}
\begin{itemize}
\item[(i)] A grayscale image has a single channel; a color image has three channels $(R,G,B)$.
\item[(ii)] Typically, each convolutional layer contains multiple kernels (filters). With $k$ filters, the number of output channels is $k$ times the number of input channels.
\item[(iii)] When the receptive fields are several times larger than the kernel, STP-CP performs a weighted aggregation over an extended region, achieving multi-scale feature extraction within a single layer while remaining padding-free. This property demonstrates the intrinsic scale adaptivity of STP-CP.
\end{itemize}
\end{rem}

\section{Cubic image\label{S7}}

This section extends the two-dimensional STP-CP framework to three-dimensional data, enabling the processing of cubic images (third-order tensors) from domains including computed tomography (CT), hyperspectral imaging, and spatiotemporal analysis.

A 3D image is a hypermatrix (or tensor, see \cite{lim13}) $A$ with entries
\begin{align}\label{sc.4.1}
A=\{a_{i,j,k}\;|\; i\in [1,m], j\in [1,n], k\in [1,\eta]\}.
\end{align}
We denote the space of such hypermatrices by $\R^{m\times n\times \eta}$.

\begin{itemize}
\item[(1)] Matrix representation of a cubic image:
\end{itemize}
Following common practice (see \cite{lim13,kil13}), a cubic image $A\in\mathbb{R}^{m\times n\times \eta}$ can be unfolded into a matrix by stacking the $\eta$ frontal slices row-wise. Specifically,
\begin{align}\label{sc.4.2} A \in \mathbb{R}^{m\times n\times \eta} \sim \mathbb{R}^{(\eta m)\times n}, \end{align} where rows are indexed by $(k, i)$ and columns by $j$.

We use an numerical example to demonstrate the design process step by step.

\begin{exa}\label{esc.4.1}

\begin{itemize}
\item[(1)] Cubic Image:
\end{itemize}

Consider a cubic image with $m=3,n=4,\eta=2$ as follows:

\begin{align}\label{sc.4.2}
A=\begin{bmatrix}
A_1\\
A_2\\
\end{bmatrix}
=\begin{bmatrix}
2&1&3&2\\
1&3&2&2\\
3&2&0&1\\
1&1&2&3\\
4&2&3&4\\
4&0&3&3\\
\end{bmatrix}\in \mathbb{R}^{6\times 4}.
\end{align}

\begin{itemize}
\item[(2)] Kernel:
\end{itemize}

A 3D kernel is likewise a cubic tensor, which is represented in matrix form as:
\begin{align}\label{sc.4.3}
K\in \R^{s\times t\times \xi}\sim \mathbb{R}^{(\xi s)\times t}.
\end{align}

Let $K\in \R^{2\times 2\times 3}$ be
\begin{align}\label{sc.4.4}
K=\begin{bmatrix}
K_1\\
K_2\\
K_3
\end{bmatrix}
=\begin{bmatrix}
1&1\\
0&1\\
1&-1\\
2&3\\
2&1\\
3&3\\
\end{bmatrix}\in \mathbb{R}^{6\times 2}.
\end{align}

\begin{itemize}
\item[(3)] Enlarged Image:
\end{itemize}

Let the enlarged (padded) image be $\bar{A}\in \R^{p\times q\times \eta}$, where
\begin{align}\label{sc.4.5}
p=m+\D m; \quad q=n+\D n,
\end{align}
where $\eta$ is unchanged and $\D m$, $\D n$ are the total increments in the vertical and horizontal sizes, respectively.
\end{exa}

\begin{exa}\label{esc.4.3} For Example \ref{esc.4.1}, let $\D m=\D n=2$. Then
$p=5$, $q=6$, $\eta=2$, and
\begin{align}\label{sc.4.6}
\bar{A}=\begin{bmatrix}
\bar{A}_1\\
\bar{A}_2\\
\end{bmatrix}=
\begin{bmatrix}
\times&\times&\times&\times&\times&\times\\
\times&2&1&3&2&\times\\
\times&1&3&2&2&\times\\
\times&3&2&0&1&\times\\
\times&\times&\times&\times&\times&\times\\
\times&\times&\times&\times&\times&\times\\
\times&1&1&2&3&\times\\
\times&4&2&3&4&\times\\
\times&4&0&3&3&\times\\
\times&\times&\times&\times&\times&\times\\
\end{bmatrix}\in \mathbb{R}^{10\times 6}.
\end{align}

\begin{itemize}
\item[(4)] RFM in the spatial plane:
\end{itemize}

Let $n_v\times n_h$ be the vertical and horizontal numbers of receptive fields; let $d_v$ and $d_h$ be the vertical and horizontal strides, and let $s\times t$ be the size of the RFM. Define
\begin{align}\label{sc.4.7}
R=P\bar{A}Q,
\end{align}
where
\begin{align}\label{sc.4.8}
\begin{array}{l}
R=P\bar{A}Q\in \mathbb{R}^{(n_v s)\times (n_h t)},\\
Q=\Xi_{n_h\times t}^{d_h}\in \mathbb{R}^{q\times (n_h t)},\\
P^{\mathrm{T}}=\Xi_{n_v\times s}^{d_v}\in \mathbb{R}^{p\times (n_v s)},\\
(n_h-1)d_h+t=q,\\
(n_v-1)d_v+s=p.
\end{array}
\end{align}
\end{exa}
\begin{exa}\label{esc.4.4} With Examples \ref{esc.4.1}-\ref{esc.4.3} and $d_h=d_v=1$,
\begin{align}\label{sc.4.9}
\begin{array}{l}
P^{\mathrm{T}}=\begin{bmatrix}
1&0&0&0&0&0&0&0\\
0&1&1&0&0&0&0&0\\
0&0&0&1&1&0&0&0\\
0&0&0&0&0&1&1&0\\
0&0&0&0&0&0&0&1\\
\end{bmatrix},\quad
Q=\begin{bmatrix}
1&0&0&0&0&0&0&0&0&0\\
0&1&1&0&0&0&0&0&0&0\\
0&0&0&1&1&0&0&0&0&0\\
0&0&0&0&0&1&1&0&0&0\\
0&0&0&0&0&0&0&1&1&0\\
0&0&0&0&0&0&0&0&0&1\\
\end{bmatrix}.\\
\end{array}
\end{align}
Then
$$
R:=\begin{bmatrix}
R_1\\
R_2\\
\end{bmatrix},
$$
where
$$
R_i=P\bar{A}_iQ,\quad i=1,2.
$$
A direct computation gives
\begin{align}\label{sc.4.10}
\begin{array}{l}
R=\begin{bmatrix}
\times&\times&\times&\times&\times&\times&\times&\times&\times&\times\\
\times&2&2&1&1&3&3&2&2&\times\\
\times&2&2&1&1&3&3&2&2&\times\\
\times&1&1&3&3&2&2&2&2&\times\\
\times&1&1&3&3&2&2&2&2&\times\\
\times&3&3&2&2&0&0&1&1&\times\\
\times&3&3&2&2&0&0&1&1&\times\\
\times&\times&\times&\times&\times&\times&\times&\times&\times&\times\\
\times&\times&\times&\times&\times&\times&\times&\times&\times&\times\\
\times&1&1&1&1&2&2&3&3&\times\\
\times&1&1&1&1&2&2&3&3&\times\\
\times&4&4&2&2&3&3&4&4&\times\\
\times&4&4&2&2&3&3&4&4&\times\\
\times&4&4&0&0&3&3&3&3&\times\\
\times&4&4&0&0&3&3&3&3&\times\\
\times&\times&\times&\times&\times&\times&\times&\times&\times&\times
\end{bmatrix}
\in \mathbb{R}^{(n_z n_v s)\times (n_h t)},
\end{array}
\end{align}
where $n_z$ is the block number in depth direction (refer to the following \eqref{sc.4.12} for details).

\begin{itemize}
\item[(5)] Masking along the depth direction:
\end{itemize}

Let
$$
R_0={\bf \times}_{(n_v s)\times (n_h t)},
$$
and define
\begin{align}\label{sc.4.11}
\bar{R}=
\begin{bmatrix}
R_0\\
R\\
R_0\\
\end{bmatrix}.
\end{align}

Let $\D \eta =2$. Then
the masked 3D image should be
$$
\bar{R}\in \R^{p\times q\times \d}
$$
where
$$
\d=\eta+\D \eta=4,
$$
Moreover, we set
$$
R_0:={\bf \times}_{8\times 10},
$$
and obtain
\begin{align}\label{sc.4.12}
\bar{R}=
\begin{bmatrix}
R_0\\
R_1\\
R_2\\
R_0\\
\end{bmatrix}.
\end{align}

\begin{itemize}
\item[(6)] RFM along the depth direction:
\end{itemize}

Let $\xi$ be the kernel depth in \eqref{sc.4.3}, stride $d_z$, and $n_z$ the number of depth receptive fields, satisfying
\begin{align}\label{sc.4.13}
(n_z-1)d_z+\xi=\d.
\end{align}
With $\d=4$, $\xi=3$, and $d_z = 1$, we have $n_z = 2$. Construct
$$
Q:=\Xi_{n_z\times \xi}^{d_z}
=\begin{bmatrix}
1&0&0&0\\
0&1&0&0\\
0&0&1&0\\
0&1&0&0\\
0&0&1&0\\
0&0&0&1\\
\end{bmatrix}.
$$
To apply this selection along depth while preserving each spatial block (of size $s\times t$) define
\begin{align}\label{sc.4.14}
H:=I_{n_v}\otimes Q\otimes I_s.
\end{align}
In this example,
$$
H=I_4\otimes Q \otimes I_2.
$$
Set
\begin{align}\label{sc.4.15}
\tilde{R}:=H\bar{R}.
\end{align}

\begin{itemize}
\item[(7)] Position rearrangement (final RFM):
\end{itemize}

To gather elements of each cubic receptive field contiguously, use a swap (perfect-shuffle) matrix:
\begin{align}\label{sc.4.16}
T=W_{[\d,n_v]}\otimes I_s.
\end{align}
Then
\begin{align}\label{sc.4.17}
C:=T\tilde{A}
\end{align}
is the final receptive-field matrix, where vertical, horizontal, and depth directions have been rearranged into block RFM-form.

The final RFM $\Psi$ is

\begin{align}\label{sc.4.18}
\Psi=
\left[\begin{smallmatrix}
\times&\times&\times&\times&\times&\times&\times&\times&\times&\times\\
\times&\times&\times&\times&\times&\times&\times&\times&\times&\times\\
\times&\times&\times&\times&\times&\times&\times&\times&\times&\times\\
\times&     2&     2&     1&     1&     3&     3&     2&     2&     \times\\
\times&\times&\times&\times&\times&\times&\times&\times&\times&\times\\
\times&     1&     1&     1&     1&     2&     2&     3&     3&     \times\\
\times&\times&\times&\times&\times&\times&\times&\times&\times&\times\\
\times&     2&     2&     1&     1&     3&     3&     2&     2&     \times\\
\times&\times&\times&\times&\times&\times&\times&\times&\times&\times\\
\times&     1&     1&     1&     1&     2&     2&     3&     3&     \times\\
\times&\times&\times&\times&\times&\times&\times&\times&\times&\times\\
\times&\times&\times&\times&\times&\times&\times&\times&\times&\times\\
\times&\times&\times&\times&\times&\times&\times&\times&\times&\times\\
\times&\times&\times&\times&\times&\times&\times&\times&\times&\times\\
\times&     2&     2&     1&     1&     3&     3&     2&     2&     \times\\
\times&     1&     1&     3&     3&     2&     2&     2&     2&     \times\\
\times&     1&     1&     1&     1&     2&     2&     3&     3&     \times\\
\times&     4&     4&     2&     2&     3&     3&     4&     4&     \times\\
\times&     2&     2&     1&     1&     3&     3&     2&     2&     \times\\
\times&     1&     1&     3&     3&     2&     2&     2&     2&     \times\\
\times&     1&     1&     1&     1&     2&     2&     3&     3&     \times\\
\times&     4&     4&     2&     2&     3&     3&     4&     4&     \times\\
\times&\times&\times&\times&\times&\times&\times&\times&\times&\times\\
\times&\times&\times&\times&\times&\times&\times&\times&\times&\times\\
\times&\times&\times&\times&\times&\times&\times&\times&\times&\times\\
\times&\times&\times&\times&\times&\times&\times&\times&\times&\times\\
\times&     1&     1&     3&     3&     2&     2&     2&     2&     \times\\
\times&     3&     3&     2&     2&     0&     0&     1&     1&     \times\\
\times&     4&     4&     2&     2&     3&     3&     4&     4&     \times\\
\times&     4&     4&     0&     0&     3&     3&     3&     3&     \times\\
\times&     1&     1&     3&     3&     2&     2&     2&     2&     \times\\
\times&     3&     3&     2&     2&     0&     0&     1&     1&     \times\\
\times&     4&     4&     2&     2&     3&     3&     4&     4&     \times\\
\times&     4&     4&     0&     0&     3&     3&     3&     3&     \times\\
\times&\times&\times&\times&\times&\times&\times&\times&\times&\times\\
\times&\times&\times&\times&\times&\times&\times&\times&\times&\times\\
\times&\times&\times&\times&\times&\times&\times&\times&\times&\times\\
\times&\times&\times&\times&\times&\times&\times&\times&\times&\times\\
\times&     3&     3&     2&     2&     0&     0&     1&     1&     \times\\
\times&\times&\times&\times&\times&\times&\times&\times&\times&\times\\
\times&     4&     4&     0&     0&     3&     3&     3&     3&     \times\\
\times&\times&\times&\times&\times&\times&\times&\times&\times&\times\\
\times&     3&     3&     2&     2&     0&     0&     1&     1&     \times\\
\times&\times&\times&\times&\times&\times&\times&\times&\times&\times\\
\times&     4&     4&     0&     0&     3&     3&     3&     3&     \times\\
\times&\times&\times&\times&\times&\times&\times&\times&\times&\times\\
\times&\times&\times&\times&\times&\times&\times&\times&\times&\times\\
\times&\times&\times&\times&\times&\times&\times&\times&\times&\times\\
\end{smallmatrix}\right].
\end{align}

\begin{itemize}
\item[(8)] CP Matrix:
\end{itemize}

Finally, partition $\Psi$ into $(n_vn_z)\times n_h$ blocks, each of size $(s\xi)\times t$. Then each block
$$
\Psi_{i,j}\in \mathbb{R}^{(s\xi)\times t},\quad i\in [1,n_vn_z],\; j\in [1,n_h],
$$
define
$$
s_{i,j}=\langle V_c(K),V_c(\Psi_{i,j})\rangle,
$$
and assemble
$$
S=(s_{i,j})\in \mathbb{R}^{(n_vn_z)\times n_h}.
$$

The final CP matrix is
\begin{align}\label{sc.4.19}
S=\frac{1}{6}\begin{bmatrix}
13&9.5&13&21.5&21\\
13&9.5&13&21.5&21\\
20&16.25&18&24.75&24.5\\
20&16.25&18&24.75&24.5\\
27.5&21.25&17.5&25&18\\
27.5&21.25&17.5&25&18\\
29.5&18&12.5&21.5&16.5\\
29.5&18&12.5&21.5&16.5\\
\end{bmatrix}.
\end{align}
\end{exa}

\section{Conclusion\label{S8}}
This paper has presented a novel convolutional neural network framework based on the semi-tensor product (STP), which replaces the conventional inner product with a dimension-free STP inner product. By leveraging this cross-dimensional operation, we have developed a padding-free convolutional product (STP-CP) that performs convolution directly over valid data regions, eliminating the need for zero-padding or explicit mask propagation. The proposed approach has been shown to effectively handle irregular images, partially damaged data, and cases with mismatched receptive field and kernel sizes, thereby reducing artifacts and noise commonly introduced by padding in traditional CNNs.

Furthermore, a unified matrix formulation for STP-based convolution has been established, demonstrating the framework’s scalability to higher-order signals such as three-dimensional volumetric data. This work provides a principled and mathematically rigorous theoretical foundation for the forward propagation of robust feature extraction in real-world scenarios involving incomplete or irregular data.

While this study has deliberately focused on the theoretical formulation and feasibility of the forward model, it opens several important avenues for future research. The immediate next steps include:

\begin{itemize}
\item {\bf Implementation of the Backward Pass and Training Analysis:} A critical next step is the formal derivation and efficient implementation of the backpropagation algorithm for the STP-CP layer. This will enable a comprehensive investigation into the training dynamics, stability, and convergence behavior of STP-CNNs, especially concerning the spatial variability of receptive field dimensions.
\item {\bf Large-Scale Empirical Validation:} Building on a trainable model, a thorough empirical evaluation is essential. This includes systematic comparisons with existing convolutional variants on public datasets to quantify gains in accuracy and robustness for tasks involving irregular data.
\item {\bf Computational Efficiency and Integration:} Future work will also focus on optimized implementations for practical deployment, a detailed analysis of computational efficiency on large-scale problems, and integration into mainstream deep-learning libraries for applications in domains like medical imaging and remote sensing.
\end{itemize}

The present work solidifies the forward propagation foundation, thereby setting the stage for these subsequent investigations into a fully trainable STP-CNN system.

\end{document}